\documentclass[10pt,conference]{IEEEtran}

\usepackage{fancyhdr}
\usepackage{tikz}
\usepackage{amsmath}

\usepackage{booktabs}
\usepackage{filecontents}

\usepackage{graphicx}
\usepackage[many]{tcolorbox}
 
\usepackage{mathtools,amssymb,latexsym,amsfonts,stmaryrd}
\usepackage{pbox}
\usepackage{amsthm}
\usepackage{multirow}
\usepackage{tikz}
\usepackage{mathrsfs}
\usepackage{mathpartir}
\usepackage{url}
\usepackage{syntax}
\usepackage{framed}
\usepackage{algorithm}
\usepackage{algorithmicx}
\usepackage[noend]{algpseudocode}
\usepackage{flushend}
\usepackage{makecell}
\usepackage{listings}
\usepackage{moresize}
\usepackage{wrapfig}
 
\usepackage{seqsplit}
\usepackage{alltt}
\usepackage{xspace}
\usepackage{enumitem}
\usepackage{pifont}

\DeclareMathAlphabet{\mathcal}{OMS}{cmsy}{m}{n}

\usepackage{amsmath, listings, amsthm, amssymb, proof, xspace}

\usepackage{thmtools}

\declaretheoremstyle[spaceabove=\topsep,notefont=\normalfont\itshape]{mystyle}

\newcommand{\revise}[2]{{\color{red}{\ifx&#1&\else- #1\fi}} {\color{ForestGreen}{\ifx&#2&\else+ #2\fi}}}%
\renewcommand{\revise}[2]{#2}%

\usetikzlibrary{arrows,patterns, decorations.pathreplacing}

\usepackage[colorlinks]{hyperref}   
\AtBeginDocument{%
  \hypersetup{pdfborder={0 0 1}, urlcolor=.}  
}

\definecolor[named]{ACMBlue}{cmyk}{1,0.1,0,0.1}
\definecolor[named]{ACMYellow}{cmyk}{0,0.16,1,0}
\definecolor[named]{ACMOrange}{cmyk}{0,0.42,1,0.01}
\definecolor[named]{ACMRed}{cmyk}{0,0.90,0.86,0}
\definecolor[named]{ACMLightBlue}{cmyk}{0.49,0.01,0,0}
\definecolor[named]{ACMGreen}{cmyk}{0.20,0,1,0.19}
\definecolor[named]{ACMPurple}{cmyk}{0.55,1,0,0.15}
\definecolor[named]{ACMDarkBlue}{cmyk}{1,0.58,0,0.21}
\hypersetup{
  colorlinks,
  linkcolor=ACMPurple,
  citecolor=ACMPurple,
  urlcolor=ACMDarkBlue,
  filecolor=ACMDarkBlue
}

\usepackage{inconsolata} 
\usepackage{biolinum} 

\usepackage{microtype}

\usepackage[noadjust]{cite}

\newcommand{\parh}[1]{\noindent\textbf{#1}}
\newcommand{\sparh}[1]{\noindent\underline{#1}}

\newcommand{\F}{Fig.}
\newcommand{\T}{Table}
\renewcommand{\S}{Sec.}
\newcommand{\A}{Alg.}

\newcommand{\ignore}[1]{}

\lstdefinestyle{base}{
  moredelim=**[is][\color{red}]{@}{@},
  escapeinside={<@}{@>}
}

\newcommand{\tool}{\textsc{CCTest}\xspace}

\usepackage{amssymb}
\usepackage{pifont}

\newcommand\DejaVuttfamily{%
  \fontfamily{DejaVuSansMono-TLF}\selectfont
}

\lstdefinestyle{base}{
  moredelim=**[is][\color{red}]{@}{@},
  escapeinside={<@}{@>}
}

\lstdefinelanguage
   [x64]{Assembler}     
   [x86masm]{Assembler} 
   {morekeywords={CDQE,CQO,CMPSQ,CMPXCHG16B,JRCXZ,LODSQ,MOVSXD, %
                  POPFQ,PUSHFQ,SCASQ,STOSQ,IRETQ,RDTSCP,SWAPGS, %
                  rax,rdx,rcx,rbx,rsi,rdi,rsp,rbp, %
                  r8,r8d,r8w,r8b,r9,r9d,r9w,r9b}} 

\renewcommand{\baselinestretch}{.99}

\usepackage{listings}
\usepackage{fancyvrb}
\usepackage{color}
\definecolor{lightgray}{rgb}{.9,.9,.9}
\definecolor{darkgray}{rgb}{.4,.4,.4}
\definecolor{purple}{rgb}{0.65, 0.12, 0.82}
\definecolor{commentgreen}{RGB}{63,127,95}

\usepackage{xcolor}
\colorlet{myPurple}{blue!40!red}
\definecolor{myOrange}{RGB}{255,192,0}
\newcommand{\code}[1]{\textcolor{black}{\textit{\bfseries{#1}}}}

 \lstdefinelanguage{Solidity}{
   keywords={len,delete,int,void,payable, public, event, contract, typeof, new, true, false, catch, function, return, null, catch, switch, var, if, in, while, do, else, case, break,unsigned,int32_t,int16_t,for},
   keywordstyle=\color{violet}\bfseries,
   ndkeywords={class, export, boolean, throw, implements, import, this},
   ndkeywordstyle=\color{darkgray}\bfseries,
   identifierstyle=\color{black},
   sensitive=false,
   comment=[l]{//},
   morecomment=[s]{/*}{*/},
   commentstyle=\color{commentgreen}\ttfamily,
   stringstyle=\color{red}\ttfamily,
   morestring=[b]',
   morestring=[b]"
 }

\newcommand{\rnum}[1]{\uppercase\expandafter{\romannumeral #1\relax}}

\usepackage{xcolor}

\algnewcommand{\LeftComment}[1]{\Statex \(\triangleright\) #1}

\definecolor{pptblue}{RGB}{194,214,236}

\begin{document}

\title{\tool: Testing and Repairing Code Completion Systems}

\IEEEoverridecommandlockouts

\author{
    \IEEEauthorblockN{Zongjie Li$^{a}$, Chaozheng Wang$^{b}$, Zhibo Liu$^{a}$, Haoxuan Wang$^{c}$, Dong Chen$^{a}$, Shuai Wang$^{*a}$\thanks{$^{*}$Corresponding author.}, Cuiyun Gao$^{b}$}
    \IEEEauthorblockA{$^a$ The Hong Kong University of Science and Technology, Hong Kong SAR}
    \IEEEauthorblockA{$^b$ Harbin Institute of Technology, Shenzhen, China}
    \IEEEauthorblockA{$^c$ Swiss Federal Institute of Technology Lausanne, Switzerland }
    \IEEEauthorblockA{\{zligo,zliudc,dchenbl,shuaiw\}@cse.ust.hk, \{wangchaozheng\}@stu.hit.edu.cn }
    \IEEEauthorblockA{\{gaocuiyun\}@hit.edu.cn, \{haoxuan.wang\}@epfl.ch }
}

\date{}

\maketitle

\begin{abstract}

Code completion, a highly valuable topic in the software development domain, has
been increasingly promoted for use by recent advances in large language models
(LLMs). To date, visible LLM-based code completion frameworks such as GitHub
Copilot and GPT are trained using deep learning over vast quantities of
unstructured text and open source code. As the paramount component and the
cornerstone in daily programming tasks, code completion has largely boosted
professionals' efficiency in building real-world software systems.

In contrast to this flourishing market, we find that code completion systems
often output suspicious results, and to date, an automated testing and
enhancement framework for code completion systems is not available. This
research proposes \tool, a framework to test and repair code completion systems
in black-box settings. \tool\ features a set of novel mutation strategies, namely
program structure-consistent (PSC) mutations, to generate mutated code
completion inputs. Then, it detects inconsistent outputs, representing possibly
erroneous cases, from all the completed code cases. Moreover, \tool\ repairs the
code completion outputs by selecting the output that mostly reflects the
``average'' appearance of all output cases, as the final output of the code
completion systems. 
With around 18K test inputs, we detected 33,540 inputs that can trigger
erroneous cases (with a true positive rate of 86\%) from eight popular LLM-based
code completion systems. With repairing, we show that the accuracy of code
completion systems is notably increased by 40\% and 67\% with respect to BLEU
score and Levenshtein edit similarity.
\end{abstract}

\IEEEpeerreviewmaketitle

\section{Introduction}
\label{sec:introduction}

Large language models (LLMs) such as GitHub Copilot~\cite{copilot}, OpenAI's
Codex~\cite{Codex}, Tabnine~\cite{Tabnine}, and Jurassic-1~\cite{jurassic} are
increasingly promoted for use within the software development domain. Machine
learning (ML) over vast quantities of unstructured text, including websites,
books, and open source code, is used to build such models, enabling them to
produce ``completions'' given inputs made up of code and comments
(documentation). To date, de facto LLM-based code completion frameworks are
advocated with the aim to provide an ``AI pair programmer'' capable of
automatically generating programs from natural language specifications or code
snippets.

Despite being a compelling and promising component in augmenting modern software
development, we observe that code completion systems are not perfect: they
frequently generate confusing and possibly erroneous results. The
``sub-optimal'' and even buggy behavior of code completion systems are
practically undesirable since it undermines the reliability and usability of
code completion. Nevertheless, it is yet neglected by today's research
community, whereas existing studies on LLM-based code completion frameworks
mainly focus on their security implication, cost reduction, or potential
extension in different
domains~\cite{zhang2022diet,pearce2022pop,pearce2021can,imai2022github,sarsa2022automatic}. 

\begin{figure}[t]
\centering
\includegraphics[width=1.01\linewidth]{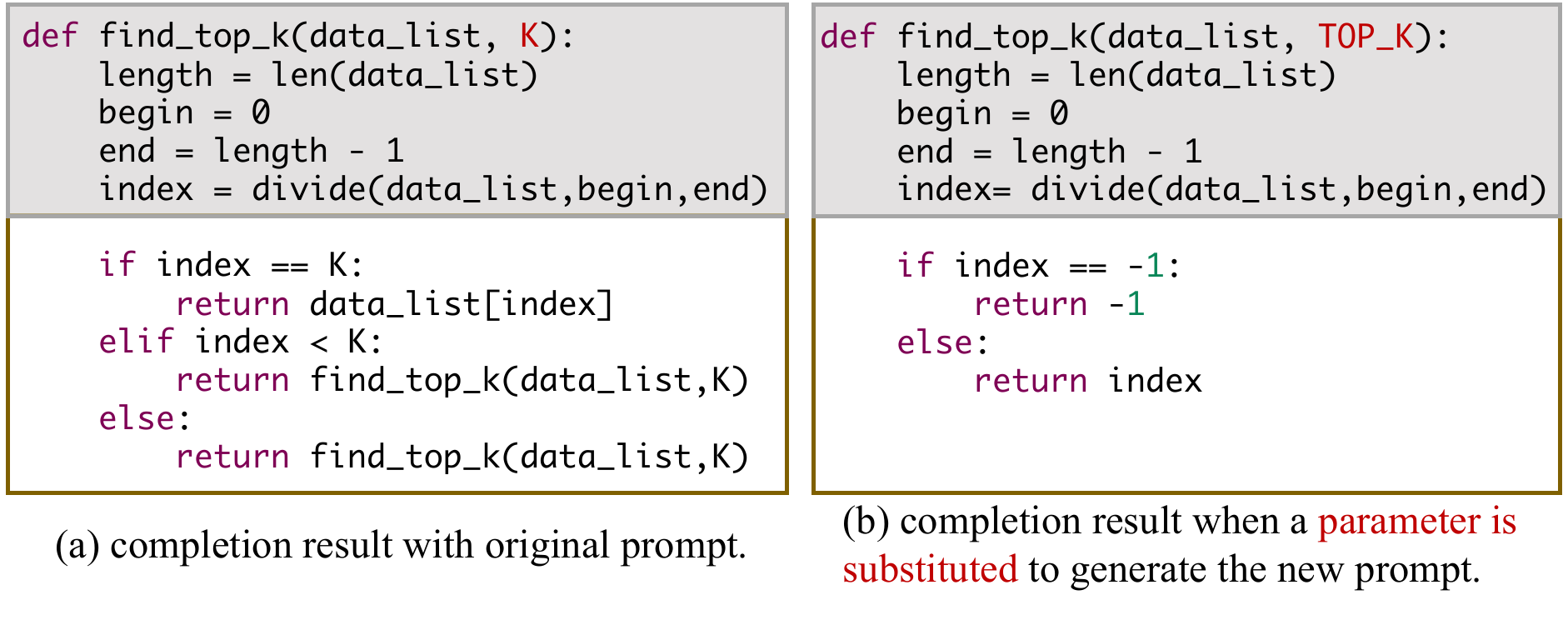}
\vspace{-25pt}
\caption{Motivating example over Codegen (input and autocompletion is marked
with gray and white background, respectively).}
\vspace{-10pt}
\label{fig:motivating}
\end{figure}

Considering \F~\hyperref[fig:motivating]{1(a)}, where a popular code completion
system, Codegen~\cite{nijkamp2022conversational}, generates a code snippet as
the completion of the input. However, by slightly tweaking the input, we observe
that Codegen outputs a dramatically different code snippet, as illustrated in
\F~\hyperref[fig:motivating]{1(b)}.
While it is generally obscure to somehow directly decide the ``correctness'' of
these two completed code snippets, given that the two inputs are identical to a
human programmer, the high distinction between two completed code snippets
indicates that Codegen's outputs are of low consistency, which is a sign of
unwanted outputs.
Overall, this apparent \textit{inconsistency} in the generated code completions
motivates us to test and enhance code completion systems. To do so, we form a
testing oracle with the intuition that the completed code snippets should be
structure-consistent.

We present \tool, an automated testing and enhancing framework for code
completion systems. Our research thrusts are two-fold. First, we design a set of
program mutation strategies, namely program structure-consistent (PSC)
mutations, to generate mutated code snippets with similar or identical program
structures.\footnote{To clarify, ``structure-consistent'' in this paper refers
to the process of transforming code snippets (i.e., inputs of code completion
systems) into mutated code snippets with identical or little altered program
structures.} Given the corresponding set $O$ of code completion outputs derived
from a seed input and its mutated inputs, we identify erroneous cases (i.e.,
outliers) in $O$ by defining and comparing distances of generated code
completions. A code completion output exhibits unusually large distances from
other outputs will be deemed spurious.
Second, We design a code enhancement strategy over completion outputs by
selecting output $\hat{o}$ that is mostly close to the ``average'' appearance of
$O$. We show that $\hat{o}$ generally manifests higher consistency with the
ground truth, extensively improving the accuracy of code completion systems.
Furthermore, \tool\ treats code completion systems as a ``black box'', so we do
not assume any specific implementation details of the code completion systems or
their underlying LLMs. 

\tool\ offers an up-to-date assessment of de facto LLM-based code completion
systems and the quality of their outputs, whose defects impede the full
potential of modern ``AI pair programmer'' in software development. 
From a total of 182,464 test inputs used for this study, we found 33,540
programs exposing code completion errors from eight widely-used LLM-based code
completion systems, one of which (Copilot) is a popular commercial tool, and the
other seven are either actively developed and maintained by the non-profit
organization (CodeParrot~\cite{codeparrot}) or hosted by the industrial
companies (EleutherAI's GPT-Neo~\cite{gpt-neo} and Salesforce's
Codegen~\cite{nijkamp2022conversational}). With enhancement, we show that the
average performance of code completion systems is notably increased by 40.25\%
and 67.43\% with respect to BLEU score and Levenshtein edit similarity.
In summary, we make the following contributions:

\begin{itemize}[noitemsep,topsep=0pt]
\item We introduce and advocate a new focus on testing and enhancing code
completion systems. This is a very important, timely topic, yet no existing
testing work has been launched. We thus envision that our efforts can guide
future research that aims to use or improve code completion tools.

\item \tool\ is an automated testing and enhancement framework for code
  completion systems even in a black-box, remote API setting. PSC
  transformations, though relatively simple, are practical and low-cost, and
  specifically designed to deliver structure-consistent mutations. \tool\ also
  features a black-box scheme to enhance the quality of completion outputs
  without the need for retraining.

  \item We evaluate one commercial (Copilot) and seven popular free code
  completion systems, in which we successfully found 33,540 programs causing
  inconsistent code completion outputs. Our enhancement further improves the
  accuracy of the code completion systems by a large margin. We made various
  observations and obtained inspiring findings regarding modern code completion
  systems. 
\end{itemize}

\parh{Artifact availability.}~We have released \tool\ to facilitate further
research~\cite{snapshot}.

\begin{figure}[!htpb]
  \centering
  \vspace{-10pt}
  \includegraphics[width=0.70\linewidth]{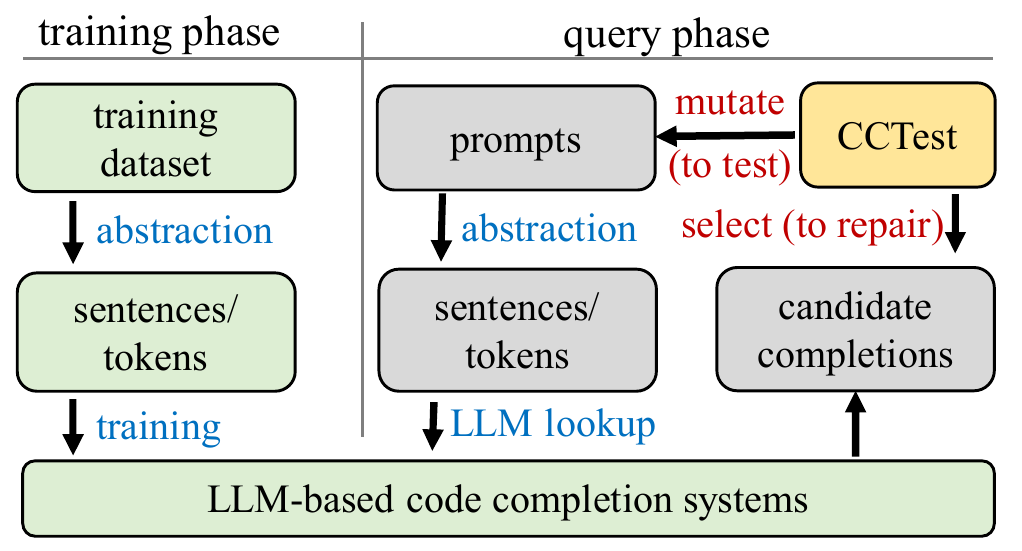}
  \vspace{-10pt}
  \caption{A holistic view of LLM-based code completion system 
  and how \tool\ facilitates testing and enhancement.}
  \vspace{-5pt}
  \label{fig:background}
\end{figure}

\section{Preliminary}
\label{sec:background}

This section introduces the background of code completion systems to deliver a
self-contained paper. Note that the automated testing and enhancement pipeline
shipped by \tool\ treats each code completion system as a \textit{black box},
which means we do \textit{not} require prior knowledge of the models or the
underlying implementation. Nevertheless, given the dominating usage of LLMs in
today's code completion
systems~\cite{copilot,Codex,DBLP:conf/emnlp/FengGTDFGS0LJZ20}, we primarily
introduce LLM-based code completion.

\F~\ref{fig:background} presents a holistic view of LLM-based code completion
systems, and explains how \tool\ fits its workflow for testing and enhancement.
Benefiting from the prosperous development and success of Transformer-based
natural language models such as OpenAI's GPT2 and
GPT3~\cite{lagler2013gpt2,chen2021evaluating,brown2020language}, the code
completion task, as a typical conditioned open-ended language generation task,
is extensively improved with much higher accuracy and applicability.\footnote{A
full introduction of transformer and its usage in open-ended language generation
is beyond the scope of this paper. Audiences can refer
to~\cite{tay2020efficient}.}

Though training data details are often obscure, modern LLMs-based code
completion systems are advertised as being trained with millions or even
billions of lines of code~\cite{copilot}. Typically, the input training data are
dissected into sentences and further into tokens, where each token comprises a
sequence of characters before being fed to LLMs. For instance, the tokenizer of
Codex and GPT-3 is configured to produce tokens with an average of four
characters. Tokens are encoded with unique numeric identifiers, up to
user-defined vocabulary size. This process is often referred to as byte pair
encoding (BPE)~\cite{gage1994new}, allowing the models to encode any rare words
with appropriate subword tokens. Various techniques are proposed to improve the
performance of LLMs~\cite{yu2022rare}, such as learning from rare tokens and
deciding on a proper stop word.

During the query phase, the code completion system's input is often referred to
as a \textit{prompt}, denoting an incomplete code snippet. Similarly, the prompt
code snippet is first abstracted into sentences and further into tokens, for
which the code completion system can predict a list of suggestions (ranked by
the confidence scores) that are more likely to continue/complete the input
prompt. For instance, the code completion system is frequently assessed by
completing a function body, given the function prototypes and some statements in
the function prologue.
Note that modern code completion systems can process prompts of different types,
including the expected program's code snippets and natural language descriptions
(comments). Such comments are usually divided into background,
input, and output to describe a competitive programming
problem~\cite{DBLP:journals/corr/abs-2203-07814}.

\tool\ is designed to test the code completion system by mutating the seed
\textit{prompts} and identifying bug-triggering prompt variants by
cross comparing the outputs. Moreover, \tool\ can enhance the code completion
output by analyzing outputs of prompt variants to select cases that are closest
to the ``average'' appearance of $O$.

\begin{figure}[!htpb]
  \centering
  \includegraphics[width=0.85\linewidth]{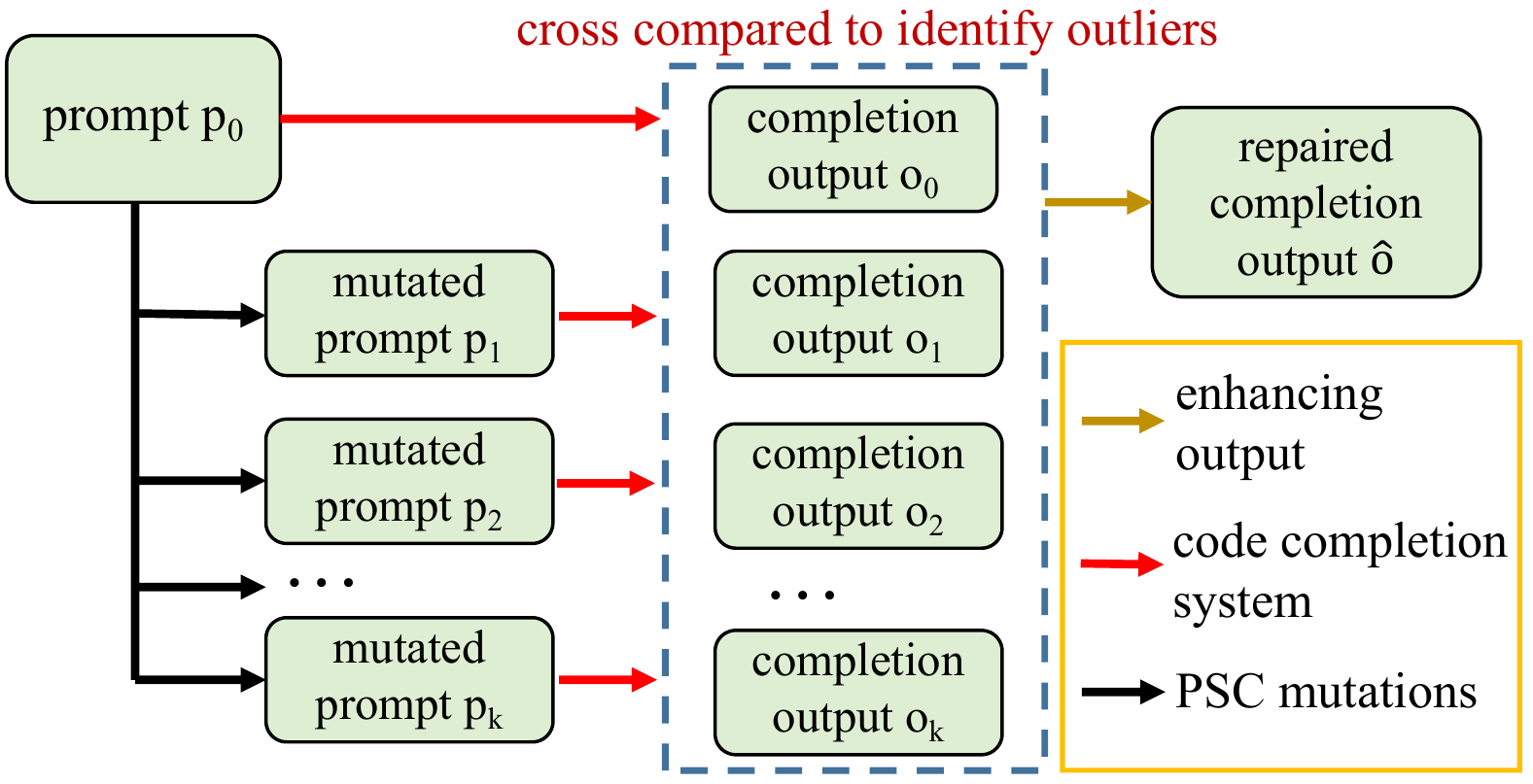}
  \vspace{-5pt}
  \caption{Workflow of \tool. \tool\ launches PSC testing by mutating a prompt
    into a collection of prompt variants, cross compares code completion
    results, and enables automated code completion enhancement.}
  \label{fig:workflow}
  \vspace{-5pt}
\end{figure}

\section{Approach Overview}
\label{sec:approach}

\F~\ref{fig:workflow} presents an overview of \tool\ in terms of testing and
enhancing code completion systems. In particular, 

\parh{\ding{172} Prompt Variants Generation.}~\tool\ launches PSC mutations and
generates a set of structure-consistent variants $P$ of an input prompt $p_0$.
Here, we propose a novel approach, PSC, that mutates a prompt with code
structure-consistent transformations. The mutated prompts manifest
closely consistent structures from human perspective. 

\parh{\ding{173} Testing Oracle Generation.}~The target code completion system
outputs a set of completion outputs in accordance with the prompt variants.
Here, we form a testing oracle over completion outputs. The key observation is
that \textit{the code completion outputs should invariably manifest high
(visual) consistency for prompt variants produced by using PSC}. Thus,
``outlier'' completion outputs are deemed erroneous, whose corresponding
prompt variants are defect-triggering inputs.

\parh{\ding{174} Completion Output Enhancement.}~The testing pipeline can be
extended to enhance code completion outputs. To this end, \tool\ identifies
output $\hat{o}$ that is closest to the ``average'' appearance of $O$ (with
respect to program distance metrics; after excluding outliers in
\ding{173}). This step does not assume any prior knowledge of the code
completion system implementation and, therefore, applies to black-box scenarios.

\parh{Study Scope.}~We primarily target the erroneous code completion outputs,
denoting ``stealthy logic bugs'' of code completion systems. As detailed in
\S~\ref{subsec:oracle}, our testing approach can automatically expose
inconsistency defects in code completion systems, meaning that when given a set
of structurally consistent prompts, the completion outputs are deemed to share
closely similar appearances as well. 

During the preliminary study, we also find that certain (mutated) prompts may
simply impede code completion systems from generating any outputs. To some
extent, this is comparable to identifying a ``crash'' of the code completion
system. Although such obviously anomalous states are not the primary focus of
\tool, we still record and report all such defects we encountered during the
evaluation.

Note that we are \textit{not} using extreme (broken) prompts to stress code
completion systems. Modern code completion models are on the basis of LLMs, and
our preliminary study shows that providing some trivial code snippets as prompts
may make the code completion system generate meaningless outputs. The current
implementation of \tool\ focuses on generating \textit{syntactically valid}
Python code snippets as testing prompts, given the Python language's popularity
and representativeness. Nevertheless, our method is \textit{not} limited to
Python; we leave supporting other programming languages as one future work.

\subsection{Program Structure-Consistent (PSC) Mutations}
\label{subsec:psc}

To test code completion systems, our key observation is that programs with
identical/mildly-changed control structures will retain such consistency in the
completion outputs. Hence, by observing certain completion outputs manifesting
high inconsistency, potential code completion errors can be flagged.

\begin{table}[!htbp]
	\centering
	\footnotesize
  \caption{Transformation schemes implemented in \tool.}
  \vspace{-5pt}
	\label{tab:psc}
	\setlength{\tabcolsep}{4.5pt}
	\resizebox{0.80\linewidth}{!}{
		\begin{tabular}{c|c|c}
			\hline
      \textbf{Class} & \textbf{Methods} & \textbf{Abbreviations} \\
	  \hline
	  \multirow{4}{*}{\makecell{Identifier \\ Level}} & rename parameter regulate & \texttt{REP_R} \\
													  & rename parameter context & \texttt{REP_C} \\
												  & rename local variable regulate & \texttt{REL_R} \\
												   & rename local variable context & \texttt{REL_C} \\
			\hline
			 \multirow{2}{*}{\makecell{Instruction \\ Level}} & instruction replacement & \texttt{IRR} \\
			  & replace bool expression & \texttt{RTF} \\
			\hline
			  \multirow{3}{*}{\makecell{Block \\ Level}} & garbage code insertion regulate & \texttt{GRA_R} \\
			   & garbage code insertion context & \texttt{GRA_C} \\
			   & print statement insertion & \texttt{INI} \\
			
			\hline
		\end{tabular}
	}
\end{table}

\parh{Design Goal.}~To generate structure-consistent mutants, we implement a set
of PSC transformations, which mutate a seed prompt from different levels of the
program hierarchy. \T~\ref{tab:psc} lists all the proposed PSC transformations.
Each of them is designed to be ``lightweight,'' in the sense that it only
slightly changes the prompts and retains the \textit{structure consistency} of
mutated prompts. However, we make an encouraging observation that such
straightforward, incremental mutations impose notable challenges for code
completion systems to process inputs and accordingly generate consistent
completion outputs. We now introduce each PSC scheme below.

\parh{\texttt{REP_R} \& \texttt{REP_C}.}~Our first two schemes rename function
parameters. Overall, they will replace one function parameter with a new
identifier, and every usage of this parameter will be identified (by
matching identifier names) and replaced accordingly. The naming can either be
``regulate'' or ``context.'' Considering the following case,

\begin{figure}[H]
  \centering
  \vspace{-10pt}
  \includegraphics[width=0.75\linewidth]{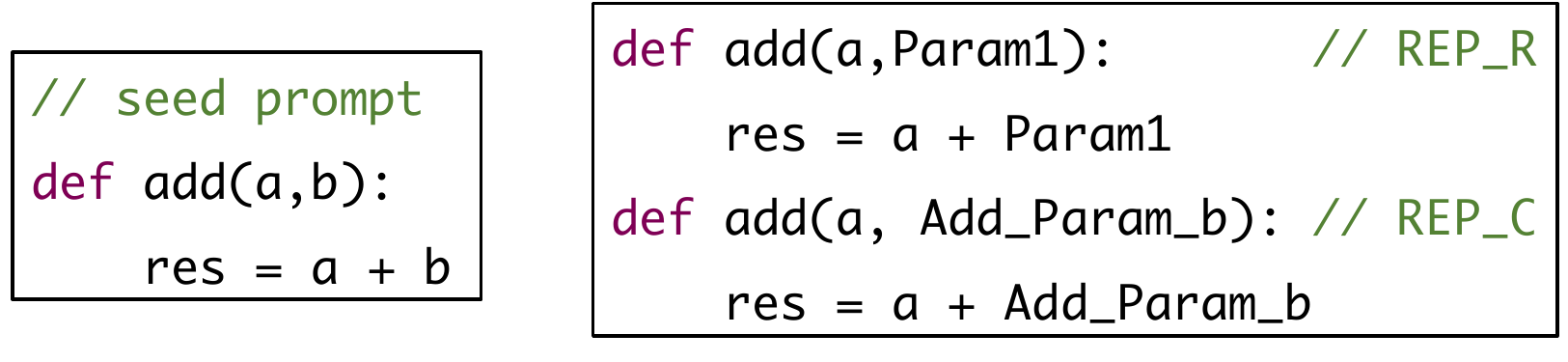}
  \vspace{-10pt}
\end{figure}

\noindent where for the ``regular'' scheme (\texttt{REP_R}), we replace the
function parameter \texttt{b} with \texttt{Param1}. As \texttt{REP_C}, which
takes the specific context into account, we extend the function parameter
\texttt{b} with a new identifier that subsumes both function name \texttt{add}
and \texttt{b}. It is evident that these two methods choose a relatively
fixed ``pattern'' to mutate prompts. However, we clarify that they are designed
by taking account developers' programming habits to replace the parameters,
which shall help (LLM-based) code completion systems avoid generating irrelevant
outputs when they behave correctly. This design intuition applies to following
PSC transformations as well, and they are very effective to provoke code
completion errors.

\parh{\texttt{REL_R} \& \texttt{REL_C}.}~These schemes rename local variables.
Similar to renaming parameters, these schemes randomly select a local variable
whose scope is in the function. Then, we replace all of its references
with a new identifier. Considering the following case,

\begin{figure}[H]
	\centering
	\vspace{-10pt}
	\includegraphics[width=0.62\linewidth]{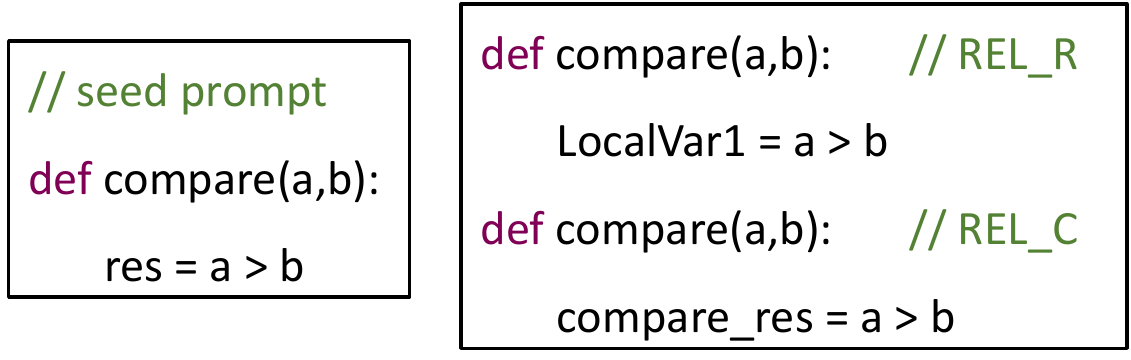}
	\vspace{-10pt}
\end{figure}

\noindent where for \texttt{REL_R}, we replace the local variable \texttt{res}
with \texttt{LocalVar1}. As for \texttt{REL_C}, which takes the context
information into account, we rewrite the local variable \texttt{res} with a new
identifier that subsumes both function name \texttt{compare} and \texttt{res}. 

\parh{\texttt{IRR}.}~This scheme implements a set of mapping rules to search and
replace certain common arithmetic operators with semantic equivalent, yet
syntactically different forms. Considering the following case, 

\begin{figure}[H]
	\centering
	\vspace{-10pt}
	\includegraphics[width=0.80\linewidth]{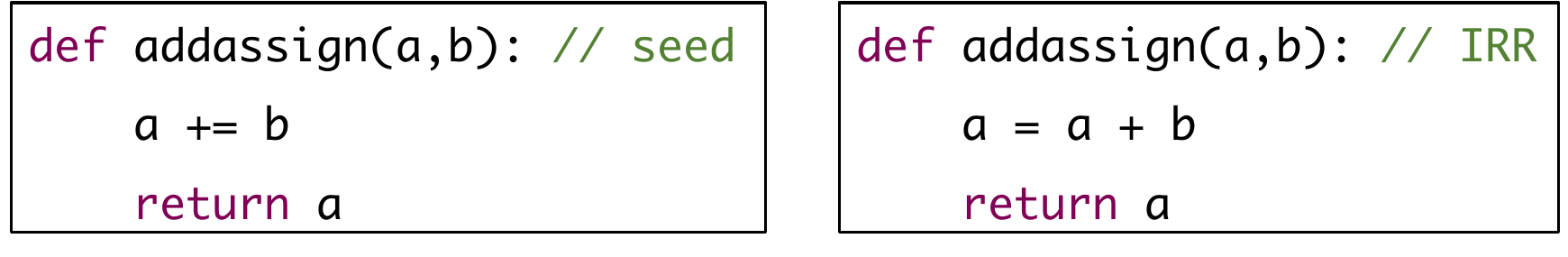}
	\vspace{-10pt}
  \end{figure}

\noindent where the \texttt{+=} operator is replaced with an addition. We
implement four mapping rules over different common arithmetic operators;
audiences can refer to our codebase for details~\cite{snapshot}.

\parh{\texttt{RTF}.}~Besides mutating arithmetic expressions, we also implement
\texttt{RTF} to mutate boolean expressions, particularly expressions used in
forming branch conditions, with their semantics equivalent variants. Considering
the following case,

\begin{figure}[H]
	\centering
	\vspace{-10pt}
	\includegraphics[width=0.87\linewidth]{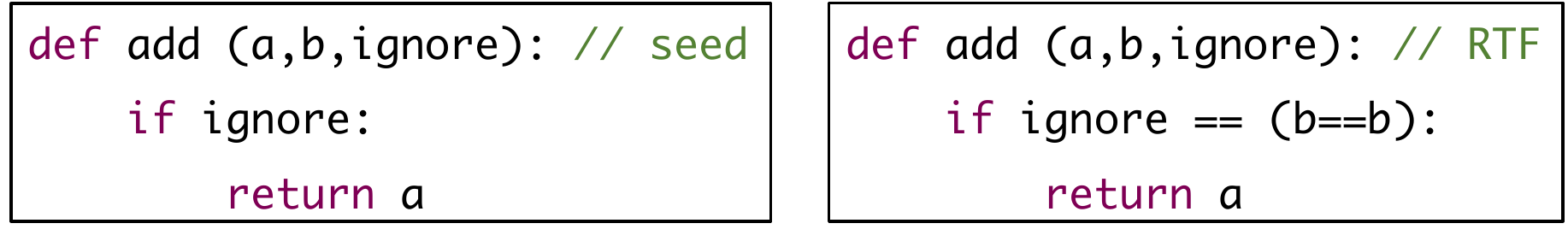}
	\vspace{-10pt}
\end{figure}

\noindent where the boolean expression is extended with an always-true
condition. This would not alter the functionality of the input prompt, nor does
it largely change the program structures. However, we find that such mutated
boolean expressions can effectively impede code completion systems from
generating structure-consistent outputs, as shown in \S~\ref{sec:evaluation}.

\parh{\texttt{GRA_R} \& \texttt{GRA_C}.}~These schemes insert a small chunk of
``garbage code'' into the program, which does not alter program semantics, but
slightly increases program control flow complexity. We use always-false
conditions to form an \texttt{if} condition, then insert a small set of
statements, which will never be executed, into the ``dead'' branch. Similar to
mutations mentioned above, creating garbage code may take context information
into account. Considering the following case,

\begin{figure}[H]
	\centering
	\vspace{-10pt}
	\includegraphics[width=0.70\linewidth]{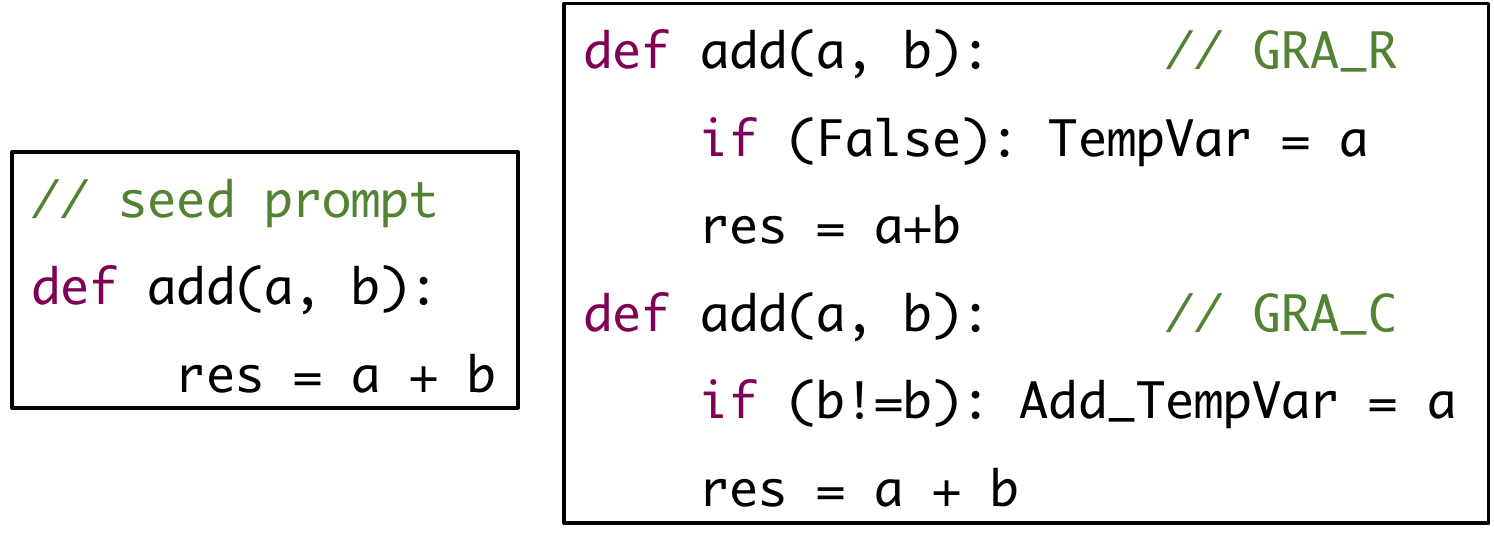}
	\vspace{-10pt}
  \end{figure}

\noindent where \texttt{GRA_R} inserts a branch whose \texttt{if} condition
expression is ``False'' with a new variable named \texttt{TempVar}. As for
\texttt{GRA_C}, which considers context (local variables and parameters), we
create an always-false condition on the basis of the syntactic form of the
function parameters, and use it to form the \texttt{if} condition expression.
Similarly, variable names in the enclosed branch also subsume both function name
and \texttt{TempVar}.

\parh{\texttt{INI}.}~The last mutation scheme inserts a \texttt{print} statement
into the prompt. This scheme is designed based on the observation that
programmers often insert such \texttt{print} statements to ease debugging. See
the following example: 

\begin{figure}[H]
	\centering
	\vspace{-10pt}
	\includegraphics[width=0.60\linewidth]{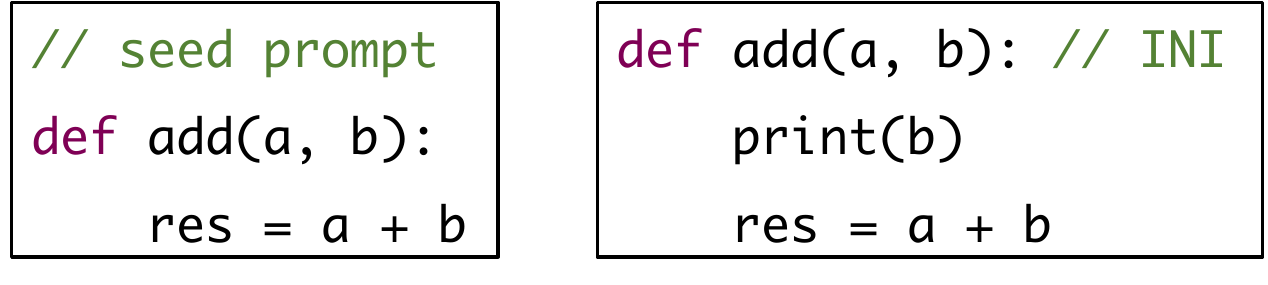}
	\vspace{-10pt}
  \end{figure}

\noindent where we insert \texttt{print} into the prompt. Though \texttt{INI}
only slightly changes a prompt, we find that its induced code completion
output effectively improves the output quality of code completion systems; see
details in \S\ref{subsec:rq3}.

\parh{Clarification.}~While applying most PSC transformations preserve the
control-flow structures of a seed prompt, garbage code insertion
(\texttt{GRA\_R} and \texttt{GRA\_C}) schemes slightly alter program structures.
Intuitively, code completion outputs for \texttt{GRA\_R}- and
\texttt{GRA\_C}-mutated prompts should appear more distinct than completion
outputs induced by other PSC schemes. However, our preliminary study and
observation show that more extensive PSC transformations may not inevitably
``dominate'' outliers in the code completion outputs. In fact, as reported in
\S~\ref{subsec:rq2}, \textit{all} PSC schemes contribute reasonably to
uncovering code completion outliers. Thus, we believe designing PSC schemes like
garbage code insertion is proper.

\parh{Alternative: Mutating Natural Language Comments.}~Besides processing
prompts in the source code form, modern code completion systems can also
generate code completions given prompts in natural language sentences. In such
cases, the input sentences are usually code comments or descriptions of the
intended code functionality. 
Careful readers may wonder about the feasibility of mutating natural language
comments to test code completion models, whose expected workflow may be similar
to recent advances in testing machine translation
systems~\cite{sun2022improving,sun2020automatic}. 
In fact, we tentatively explored this direction. We clarify that unlike machine
translation system testing, whose inputs are \textit{arbitrary} natural language
sentences, code comments often have \textit{limited mutation space}; for
instance, a vast majority of code comments have no
adjectives~\cite{DBLP:journals/corr/abs-2202-01142,DBLP:journals/corr/abs-2112-02125}.
This distinction makes it hard to mimic testing methods for machine
translation.

\subsection{PSC Testing: Forming Testing Oracles}
\label{subsec:oracle}

\begin{algorithm}[!htpb]
	\footnotesize
	\caption{Outlier selection algorithm.}
	\label{algo}
	\begin{algorithmic}[1]
	\Statex \textbf{Input:} $\mathcal{O}$: Code completion output set of size $k$ 
	\Statex \textbf{Input:} $T$: threshold
	\Statex \textbf{Output:} $\mathcal{L}$: Outliers
	\State $ScoreMatrix$ = []
	\For{$i$ in 1 to k}
		\For{$j$ in i to k}
		 $ScoreMatrix[i][j] =  \code{Sim}(o_i, o_j)$\
		\EndFor
	\EndFor
	\State $\code{Normalize}(ScoreMatrix)$
	\For{$i$ in 1 to k}
	\State count = 0
	\For{$j$ in 1 to k}
	\If{$ScoreMatrix[i][j] < \code{Median}(ScoreMatrix) $}
	 \State count = count + 1
		\If{$\text{count} \ge T$}
		\State $\mathcal{L}.append(o_{i})$
		\State \textbf{break}
		\EndIf
	\EndIf
	\EndFor
	\EndFor
	\State \Return $\mathcal{L}$
	\end{algorithmic}
\end{algorithm}

Given a list of code completion outputs $\mathcal{O}$ in accordance with the
mutated prompts $\mathcal{P}$, we compare each output $o_{i} \in
\mathcal{O}$ with the remaining cases in $\mathcal{O}$ and decide if it is an
``outlier.'' The complete algorithm is shown in Algorithm~\ref{algo}. 
We first iterate each case in $\mathcal{O}$, decide its similarity with the
remaining cases (lines 2--3), and normalize the scores (line 4). Here,
\code{Sim} is implemented using the Levenshtein string-level edit similarity
provided by fuzzywuzzy~\cite{Fuzzywuzzy}, a standard algorithm to decide the
edit similarity among two programs.
Then, we employ a threshold $T$ to decide whether a case exhibits anomalously
low similarity with other cases for more than $T$ times (lines 5--12) and deem
the abnormal case as an outlier. The corresponding $p_{i}$ is deemed as an
error-inducing input. 
Overall, given that we have implemented $N$ ($N=9$ in the current implementation
of \tool) schemes to mutate a prompt, $T$ is a configurable hyperparameter ($T
\le N$), such that a code completion output deems an outlier if its edit
distance scores with $T$ code completion outputs are less than the average
distance score (computed by \code{Median}) of code completion output pairs.

$T$ will be decided with empirical evidence, as will be discussed in
\S~\ref{subsec:rq2}. After performing Algorithm~\ref{algo}, we keep the
remaining code completion outputs $\mathcal{O}^{*} = \mathcal{O} \setminus
\mathcal{L}$ for usage in the enhancement phase (\S~\ref{subsec:repairing}).

\parh{Alternative: \code{Sim} Metrics.}~One may question the current
implementation of \code{Sim}, which is based on string-level edit distance
rather than structure-level distance. During preliminary study,
we explored using structure-level distance metrics to implement \code{Sim}.
Nevertheless, speed is a main bottleneck, as tree-edit distance is much slower.
Moreover, code completion outputs are usually code snippets consisting of
approximately ten (or fewer) lines of code whose ``structures'' may be less
significant; we find that standard tree-edit distance metrics often give vague
results. Thus, we deem that using Levenshtein string-level edit distance proper
and practicable for our testing.

\begin{figure}[!htbp]
  \vspace{-5pt}
  \centering 
  \includegraphics[width=1.00\linewidth]{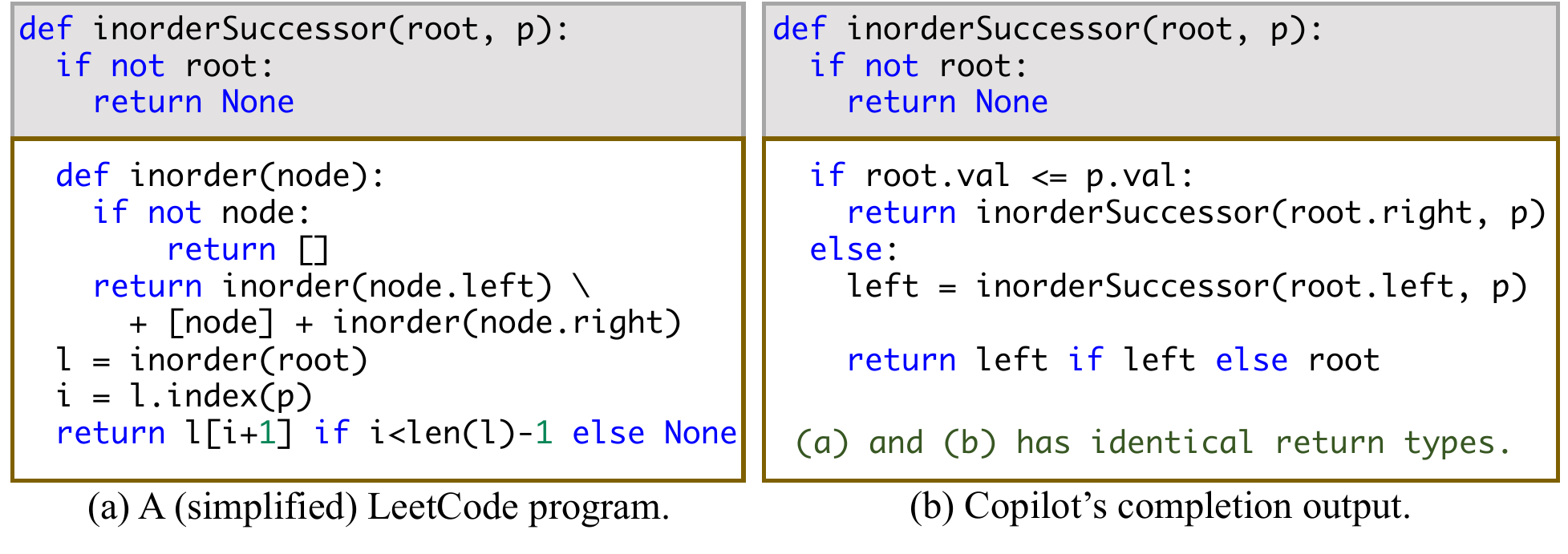}
  \vspace{-20pt}
  \caption{Comparing the syntactic form with ground truth appears overly strict
  and lead to false positives. The code snippets are simplified for
  readability.}
  \label{fig:comparison}
\end{figure}

\parh{Alternative Testing Oracles.}~We deem our formed
structure-consistency testing oracle as an instance of metamorphic testing
oracle~\cite{chen1998metamorphic}.
One may wonder about the feasibility of forming alternative testing oracles.
Below, we discuss two alternative oracles and deliberate their (in-)capability
in our research.

\sparh{Comparing Syntactic Form with Ground Truth.}~One way of forming an oracle
is to cut a program $p$ into two chunks $p_1$ and $p_2$, whereas $p_1$ serves
the prompt, and $p_2$ serves the ground truth. We test code completion by
comparing the syntactic-level equality between $o_{1}$ (completion output of
$p_1$) with $p_2$.

We clarify that this oracle is often \textit{too strict}.
\F~\ref{fig:comparison} compares the code completion output of Copilot and the
ground truth code snippet (this code snippet is from LeetCode
\cite{Leetcode0285}). While they have identical semantics, their syntactic forms
are distinct. The ground truth uses a helper function \texttt{inorder} to
deliver a classic inorder traversal of the input binary tree, whereas Copilot
output appears to be leveraging recursive calls. Overall, though the code
completion output may frequently appear syntactically distinct from ground
truths, the completion outputs should be deemed as functionally correct.
Therefore, comparing the syntactic equality of completion outputs with ground
truth may not faithfully reflect true defects of code completion systems. Such a
strict oracle may induce many false positives.

\sparh{Checking Functionality According to Ground Truth.}~Recent works explored
strictly checking the functionality correctness of code completion outputs in
accordance with ground truth.~\cite{DBLP:conf/msr/NguyenN22} tests Copilot using
prompts created from LeetCode programs, then counts the number of passed
LeetCode test cases over the code completion outputs (each completion output
forms a LeetCode solution, when put together with the prompts). Similarly,
HumanEval~\cite{chen2021codex} assesses the functionality correctness of code
completion outputs with hand-written programming problems. Overall, for those
approaches, more passed test cases indicate that the completion outputs are of
higher quality. 
Those approaches, however, may be less desirable than our formed oracle. Modern
code completion systems are \textit{not} championed to replace human
programmers; rather, it aids human programmers by suggesting useful code
snippets. Therefore, strictly correct functionality may not be the first design
target for a code completion system, e.g., some sloppy arithmetic errors in
completion outputs may be easily fixed by users. Thus, we believe it is more
desirable to cross compare the structural-level consistency (as in \tool),
rather than using test inputs to check the functionality correctness.

\subsection{Completion Output Enhancement}
\label{subsec:repairing}

This section presents the technical pipeline of enhancing the code completion
outputs. By launching the testing in \S~\ref{subsec:oracle}, we collected a set
of code completion outputs $\mathcal{O}^{*}$, with outliers being excluded. 
After that, we aim to identify a code completion output that appears mostly
close to the ``average appearance'' of $O^{*}$. To this end, we measure the
average pair-wise edit similarity $\hat{s}$ of every $o_i, o_j \in O^{*}$. Note
that at this step, we re-use the \code{Sim} function implemented in
\A~\ref{algo}, meaning that we compute the Levenshtein string-level edit
distance as the similarity score.
Then, we search for an $\hat{o} \in O^{*}$, whose average pair-wise similarity
scores with all other elements in $O^{*}$ is the closest to $\hat{s}$. This
$\hat{o}$ will be the repaired code completion output returned to the users.

To clarify, \tool\ aims to enhance the quality of code completion outputs, in
the sense that the repaired output is closer to the ``average-looking'' output.
As in our evaluation (\textbf{RQ3}; see \S~\ref{subsec:rq3}), we use edit
similarity and BLEU scores as metrics to unveil that \tool\ can find better
completions, by showing that the repaired outputs have closer distance with the
ground truth under both metrics.
Nevertheless, \tool\ cannot guarantee that the repaired outputs become
``semantically correct'', if the original outputs are not. In other words,
\tool\ does \textit{not} directly repair the semantics-level
defects/inconsistency. Semantics-level code repairing is inherently hard. More
importantly, it is unaligned with the design goal of \tool. As already noted in
\textbf{Alternative Oracle} above, we believe it is less proper to check the
functionality correctness in this research context.

\section{Implementation and Evaluation Setup}
\label{subsec:study-setup}

\tool\ is implemented in Python, with about 5k LOC. \tool\ currently focuses on
mutating Python code, given the popularity of this language in software
development and the matureness of corresponding code completion systems. We now
discuss the implementation and evaluation setup. 

\parh{Parsing and Mutating Programs.}~We parse Python code with
tree-sitter~\cite{treesitter}, a mature and dependency-free parser generator
tool with an open-source license. It is widely used in code-related projects
such as Codebert~\cite{DBLP:conf/emnlp/FengGTDFGS0LJZ20}, and it does not
require the input code to be executable, meaning that incomplete code fragments
without a building script can also be parsed. We first parse the prompt code
into a concrete syntax tree, and conduct several sanity checks (see in
\S~\ref{subsec:rq1}) on the target code to see which PSC transformation could be
performed. Then, after applying feasible PSC transformations, \tool\ will output
the corresponding transformed prompts with IDs of the applied PSC transformations.

\begin{table}[!thp]
	\centering
	\scriptsize
  \vspace{-5pt}
	\caption{Statistics of the Python prompts used in the study.}
  \vspace{-5pt}
	\label{tab:statistics}
  \resizebox{0.90\linewidth}{!}{
  \begin{tabular}{l|c|c}
  \hline
   \textbf{Split} & LeetCode &CodeSearchNet \\
   \hline
   Total \#  seed programs &613  & 2,297\\
   Average \#  token per seed & 149.4 & 115.4 \\
   Max \#  token among seeds   & 513   &  623 \\
	Total \#  generated variants &4,296  & 15,602\\
	Average \#  token per mutated seed & 151.6 & 117.3 \\
  Max \#  token among mutated seeds     & 525   &  629 \\
  \hline
	\end{tabular}
  }
  \vspace{-4pt}
  \end{table}

\parh{Seed Programs.}~Consistent with most research in this field, we form our
evaluation dataset from a popular repo of LeetCode solution
programs~\cite{leetcode} and CodesearchNet~\cite{husain2019codesearchnet}.
LeetCode is an online platform to practice algorithmic coding challenges for
technical interviews. As will be noted in \S~\ref{tab:tool}, since LLM-based
code completion systems limit the max token size, we only select programs whose
token length is between 32 and 2048. Overall, each seed program contains a
medium size function with generally complex control structures. As in
\T~\ref{tab:statistics}, we pick 613 code snippets from the LeetCode solution
repo as the seed programs. The total number of generated variants for LeetCode
programs is 4,296 (recall not every PSC transformation is applicable to
arbitrary Python code). 

CodeSearchNet is particularly designed in the research of code representation
learning. The Python component of this dataset contains train, validation, and
test splits; we use the test split for the evaluation. Similar to LeetCode, we
only select programs with token length between 32 and 2048, and for the code
snippets which share the same attribute named ``path'', we only maintain one to keep
the result balanced.

We clarify that both LeetCode and CodeSearchNet are deemed proper for
neural code learning tasks like code completion~\cite{husain2019codesearchnet}
and they are extensively used in benchmarking relevant models~\cite{guo2022unixcoder,guo2020graphcodebert,DBLP:conf/emnlp/FengGTDFGS0LJZ20,DBLP:conf/emnlp/0034WJH21}.
Thus, errors and enhancement demonstrated in our evaluation should mostly reflect
common obstacles and improvement users can expect in real-life scenarios. In
contrast, \textit{overly complex prompts} (e.g., real-world complex software)
may unavoidably impede the understanding and assessment of LLM-based code
completion systems.

\parh{Statistics of Test Cases.}~\T~\ref{tab:statistics} reports statistics of
the test cases. We collect a total of 2,910 programs from LeetCode/CodeSearchNet
as test seeds for each code completion system. The total number of generated
variants is 19,898 (see \T~\ref{tab:pass-distribution} for the breakdown). For
each seed program with its variants, we equally split the function into
two parts: the first part is used as a prompt, and the remaining is used as
``ground truth'' to assess our enhancement, see details in \S~\ref{subsec:rq3}.

\begin{table}[t]
	\centering
	\scriptsize
  \caption{Code completion systems evaluated in the study.}
  \vspace{-5pt}
	\label{tab:tool}
	\resizebox{0.67\linewidth}{!}{
		\begin{tabular}{l|c|c|c}
			\hline
      \textbf{System} & \textbf{\#}   & \textbf{\# Vocab} & \textbf{\# Max.} \\
      \textbf{Name} & \textbf{Params} & \textbf{(tokens)} & \textbf{tokens}  \\
		
      \hline
            Copilot          & ? & ? & ?  \\
            CodeParrot-small & 110M & 32,768 & 1024  \\
            CodeParrot       & 1.5B & 32,768 & 1024\\
            GPT-Neo-125M     & 125M & 50,257 & 2048  \\
            GPT-Neo-13B      & 1.3B & 50,257 & 2048  \\
            GPT-J            & 6B & 50,400 & 2048  \\
            Codegen-2B       & 2B & 50,400 & 2048  \\
            Codegen-6B       & 6B & 50,400 & 2048  \\
			\hline
		\end{tabular}
	}
  \vspace{-5pt}
\end{table}

\parh{Code Completion Systems.}~\T~\ref{tab:tool} reports the code completion
systems used in the evaluation.
First, we use Github Copilot, one highly visible commercial code completion
system that generates quite a buzz in the community~\cite{twcopilot}. We
purchase the standard commercial license (for single user) to unleash its full
potential. As a ``black-box'' commercial product, it is unclear about
their implementation (marked as $?$ in \T~\ref{tab:tool}). 

We also evaluate several well-known LLM models for code completion, including
CodeParrot~\cite{codeparrot}, GPT-Neo~\cite{gpt-neo}, GPT-J~\cite{gpt-j}, and
CodeGen~\cite{nijkamp2022conversational}. All of these models are deemed to
employ large-scale language models, given that up to billions of parameters are
involved in their underlying models, and tens of thousands of vocabularies are
considered. CodeParrot is a GPT-2 model trained specifically for Python code
generation. We use two variants, CodeParrot-small and CodeParrot. The smaller
variant contains less amount of parameters and is trained over fewer data.
Nevertheless, both variants manifest a high level of code generation capability.
As for GPT-Neo, we use two variants, GPT-Neo-125M and GPT-Neo-13B, which are
both GPT3-like models. Our observation shows that GPT-Neo-125M is prone to
generate less diverse but more straightforward code snippets compared to its
larger variant. Given that said, we believe both models are well-suited for code
generation and manifest reasonably high robustness under our testing campaign.
GPT-J, also referred to as GPT-J-6B, is a transformer model with 6B parameters
in the pre-trained model. Two versions of CodeGen specially designed for program
synthesis are evaluated. Both versions, CodeGen-2B-mono and CodeGen-6B-mono, are
trained on a large corpus of Python code. 

All these systems are stated to be trained on a large corpus of open-source
source code. For instance, Copilot is noted to include the vast majority of
GitHub's open-source code. GPT-J and GPT-Neo are trained on the Pile
dataset~\cite{pile}, a diverse language modeling dataset.
We download their pre-trained models from huggingface~\cite{huggingface} and run
the code completion locally for all LLMs except Copilot.
LLMs support different search (sampling) strategies to generate code
completions, such as beam search, temperature sampling. These factors make our
testing pipeline less deterministic. Thus, in our experiments, we follow
Copilot's interface to only choose the completion result with the highest
confidence score as its ``output.'' For other tested models, we disable their
sampling strategy.

In all, LLM-based code completion systems presented in \T~\ref{tab:tool}, to the
best of our knowledge and experience, represent the best systems available to
the public. We tentatively explored code completion solutions based on
conventional machine learning or rule-based approaches. We clarify that these
conventional methods were seen to produce much worse and shallow code completion
outputs compared with these LLM-based modern systems.

\section{Findings}
\label{sec:evaluation}

In evaluation, we mainly explore the following research questions.
\textbf{RQ1}: Can \tool\ generate high-quality and structure-consistent prompt variants? 
\textbf{RQ2}: How effective is \tool\ in detecting code completion defects?
\textbf{RQ3}: To what extent can \tool\ enhance the quality of code completion outputs?
We answer these RQs in subsections below.

\subsection{RQ1: Effectiveness on Input Generation}
\label{subsec:rq1}

Answering \textbf{RQ1} requires assessing the quality of mutated prompts. At
this step, for each seed prompt, we generate mutants and first check whether
they pass the parsing. In particular, we generate up to 9 mutants for each
prompt, leading to a total of 19,898 mutant prompts generated on top of 2,910
seed prompts.  We use the standard \texttt{ast} module in Python to parse all transformed Python
prompts into abstract syntax trees to check their validity.
All generated mutant prompts are parsed without any 
error, indicating that they are grammatically valid.

\begin{table}[!thp]
  \centering
  \vspace{-5pt}
  \scriptsize
  \caption{Distribution of structural consistency scores.}
  \vspace{-5pt}
  \label{tab:distribution}
  \setlength{\tabcolsep}{1.5pt}
  \resizebox{1.00\linewidth}{!}{
    \begin{tabular}{l|c|c|c|c|c|c}
      \hline
      \textbf{Distance} & $[0,0.05]$ & $ [0.05,0.1]$ & $[0.1,0.15]$ & $[0.15,0.2]$ & $[0.2,0.9]$ & $[0.9,1.0]$ \\
      \hline
      \textbf{Freq. (\%)}             & 90.16   & 8.30  & 0.92   &0.30 & 0.32 & 0 \\
      \textbf{Cumulative Freq. (\%)}  & 90.16   & 98.46  & 99.38 &99.68 & 100.0 & 100.0 \\
      \hline
    \end{tabular}
  }
  \vspace{-5pt}
\end{table}

To illustrate the structure consistency of mutated prompts, we use
\texttt{pycode-similar}~\cite{pycodesim}, a well-performing tool to calculate
Python code similarity. This similarity metric is based 
on AST structures, which is different from Levenshtein
edit similarity that we use in deciding the ``outliers''. Let
the prompt AST have $n$ nodes, where $m$ nodes are matched toward nodes on the
mutated prompt's AST. \texttt{pycode-similar} returns ratio $\frac{m}{n}$,
denoting how similar two programs are. We report the distribution of the
``distance score'' (distance is computed as $1 - \frac{m}{n}$) in
\T~\ref{tab:distribution}. For the vast majority of
mutated prompts (over 98\%), structural-level distance is less than 0.1, meaning
over 90\% of AST nodes are matched.
Recall that several PSC transformations (\S~\ref{subsec:psc}) introduce only
identifier-level changes. As expected, the distance scores between their mutated
and seed prompts are zero, implying that program structures are retained.
A few cases have slightly higher distances. With manual inspection, we find that
it is primarily because the size of original prompts is short, leading to an
increased distance $1 - \frac{m}{n}$ where $n$ is small.

\begin{tcolorbox}[size=small]
\textbf{Answer to RQ1}: \tool\ can generate large-scale, grammatically valid,
and structurally consistent prompt mutants.
\end{tcolorbox}

\subsection{RQ2: Bug Detection}
\label{subsec:rq2}

\begin{table*}[!thtp]
  \centering
  \caption{Overview of outlier detection results. For each system, we use 4909
  LeetCode prompts and 17899 CodeSearchNet prompts to test.}
  \vspace{-8pt}
  \label{tab:results}
  \setlength{\tabcolsep}{2pt}
  \resizebox{0.65\linewidth}{!}{
    \begin{tabular}{l|c|c|c|c|c|c}
      \hline
      \multirow{2}{*}{\textbf{System}} & \multirow{2}{*}{\textbf{\#No Results}} & \multicolumn{5}{c}{\textbf{\#Outliers}} \\\cline{3-7}
                                       &                                         &T= 1 &T= 3 &T= 5 &T= 7 &T= 9 \\
      \hline
      Copilot           & 4+37  & 3003 + 12101 & 1347 + 7928 & 803 + 5570 & 559 + 3899 & 293 + 2184 \\
      \hline                                                                                                                          
      CodeParrot        & 1+0  & 4798 + 17605 & 4379 + 16118 & 3631 + 13368 & 2359 + 9023 & 904 + 3778 \\
      \hline                                                                                                            
      CodeParrot-small  & 2+0 & 4812 + 17611 & 4469 + 16069 & 3776 + 13454 & 2470 + 9344 & 1033 + 4009 \\
      \hline
      GPT-J             & 0+0& 4606 + 17280 & 3776 + 15073 & 2832 + 12005 & 1786 + 7875 & 586 + 3174 \\
      \hline
      GPT-NEO-13B       & 1+0 & 4729 + 17427 & 4088 + 15495 & 3311 + 12536 & 2120 + 8462 & 794 + 3463 \\
      \hline
      GPT-NEO-125M      & 0+2& 4734 + 17509 & 4281 + 15556 & 3654 + 12907 & 2523 + 8966 & 1079 + 3700 \\
      \hline
      Codegen-2B-mono   & 2+9 & 4661 + 17221 & 3794 + 15112 & 2731 + 12241 & 1638 + 8460 & 639 + 3761 \\
      \hline
      Codegen-6B-mono   & 0+0   & 4578 + 17016 & 3575 + 14595 & 2493 + 11546 & 1452 + 7866 & 584 + 3559 \\
      \hline
      \hline
      Total             & 10+48  & 35921 + 133770 & 29709 + 115946 & 23231 + 93627 & 14907 + 63895 & 5912 + 27628\\
      \hline
    \end{tabular}
  }
  \vspace{-8pt}
\end{table*}

To answer \textbf{RQ2}, we first launch testing to detect code completion
defects. \T~\ref{tab:results} reports our findings. Note that we use two
datasets for testing, including 4,909 prompts from the LeetCode dataset
and 17,899 prompts from the CodeSearchNet dataset. In
\T~\ref{tab:results}, ``3003 + 12101'' in the first ``\#Outliers $T=1$'' cell
means that 3,003 outliers are found using the LeetCode mutated inputs, whereas
12,101 outliers are found using the CodeSearchNet mutated inputs. The same
format applies for the ``\#No Results'' column.

First, as shown in the second column of \T~\ref{tab:results}, while nearly all
mutated prompts can be processed, we still find 58 (0.03\%) mutated
prompts that trigger ``no response'' for the tested code completion systems. As
expected, such cases are rare, given the high capability and comprehensiveness
of LLM-based production code completion systems.

For the outlier detection evaluation, we assess the performance under different
values of the hyper-parameter $T$. As clarified in \S~\ref{subsec:oracle},
\tool\ leverages a threshold $T$ to form the testing oracle: a code completion
output is deemed an outlier if it has a small similarity score with $T$ code completion
outputs mutated from the same seed prompt. Therefore, $T$ is an integer ranging
from 1 to the total number of schemes ($9$ in our implementation).
\T~\ref{tab:results} reports the number of uncovered outliers in accordance with
different models and thresholds. Overall, out of 182,464 test cases, a
substantial number of defects are found under all five configurations. 
As expected, the increment of $T$ decreases the number of detected defects
(outliers). As we illustrated in Algorithm~\ref{algo}, the selection stringency
for outliers depends on the threshold $T$, where a higher $T$ represents a
stricter standard. Also, different from some robustness testing papers
(e.g.,~\cite{goel2021robustness}) that merely detecting mutated inputs that lead
to outlier predictions, bad completions are derived from either original (what
users start with) or mutated prompts in this study. For instance, when $T=9$,
among all 33,540 (5912 + 27628) bad completions, 3,008 ($8.9\%$) are outputs
derived from the original prompts. That is, a reasonable portion of ``bad
completions'' are from original prompts.

Copilot outperforms the other seven LLMs, given fewer inconsistency defects
(293+2184 for $T = 9$). However, Copilot has the most ``no response'' failures
(41 out of 58). We suspect that Copilot will refuse to return any output when
its model in the remote server fails to generate code snippets with valid syntax
or high confidence scores.
Despite the small number of ``no response'' failures, all the other cases can be
processed by code completion systems to produce non-trivial outputs.

\begin{table}[!thp]
  \centering
  \scriptsize
  \caption{Assessing \tool's findings with manual investigation.}
  \vspace{-5pt}
  \label{tab:rq2-statistics}
  \resizebox{0.72\linewidth}{!}{
    \begin{tabular}{l|c|c|c|c|c} \hline 
              & T=1    & T=3    & T=5    & T=7    & T=9    \\ \hline
    TP        & 216    & 342    & 429    & 537    & 689   \\
    FN        & 26     & 38     & 70     & 90     & 104    \\
    Precision & 0.270  & 0.427   & 0.536 & 0.671 & 0.861 \\
    Recall    & 0.892  & 0.900   & 0.859 & 0.856 & 0.868 \\
    F1 score  & 0.414  & 0.579   & 0.660 & 0.752 & 0.865 \\ 
    \hline
    \end{tabular}
  }
  \vspace{-8pt}
\end{table}

\noindent \textbf{Manual Validation.}~At this step, we first measure the true
positive (TP) rate of the outliers found under different thresholds $T$. For
each pair of $\langle T,model \rangle$, we randomly sample 100 cases from two
datasets, resulting in a total of 4,000 ($5 \times 8 \times 100$) cases. The
first two authors check each case to manually decide if an outlier is TP or
false positive (FP).
Our manual inspection results are presented in the ``TP'' row of
\T~\ref{tab:rq2-statistics}. It shows that with the increment number of $T$, the
number of TPs keeps increasing. Particularly, when $T=9$, out of 800 positive
findings of \tool, we have only 111 ($800-689$) cases that are FPs. To better
understand our findings, we further analyze those remaining 111 cases and
summarize the following two main reasons for FPs. 

$(i)$ The completion results are highly robust for the majority of mutants. To
select an outlier, Algorithm \ref{algo} will cross compare the outputs from the
same seed program. However, it would be possible that all other results are
exactly the same except for one mutant, which would be treated as an
``outlier'', even if it is only slightly different from the others. Such cases
count for the 59.46\% of FPs. $(ii)$ No ``mainstream'' results. The completion
results for some vague prompts would vary drastically, and sometimes the vast
majority, if not all completion outputs, appear to be different. During the
manual inspection of such cases, authors would not deem any mutants as
``outliers.'' Such cases count for 26.13\% of FPs.

Recall that our test oracle uses string-level distance metrics to decide
outliers, instead of directly using program structure-level distance metrics. We
have clarified why string-level distance is preferable and practical to
structure-level distance in \textbf{Alternative \code{Sim} Metrics} paragraph of
\S~\ref{subsec:psc}. From the above manual inspection, it is evident that using
string-level distance metrics does not primarily add FPs; we find that when two
completion outputs exhibit a high distance in string-level metrics, they
generally look distinct from program structures as well. Overall, empirical
results here further support using string-level distance metrics in deciding
outliers.

The above study explores TP/FP over 4,000 ``positive'' findings. As a testing
tool, \tool\ cannot avoid false negatives (FNs). Nevertheless, measuring FNs
help understand the potential of \tool. Hence, at this step, we randomly select
4,000 negative samples from \tool's findings. Two authors manually check the FN
and true negative (TN) cases (the procedure follows how we confirm TP/FP). We
report the number of FNs in \T~\ref{tab:rq2-statistics}. We accordingly compute
the precision, recall, and F1 scores. It is observed from the ``FN'' row of
\T~\ref{tab:rq2-statistics} that the number of FNs increases when $T$ grows.
This is reasonable, given that the higher $T$ is, the more likely \tool\ may
miss some findings. Nevertheless, \tool\ becomes more accurate when $T$ grows,
as reflected by the F1 scores in \T~\ref{tab:rq2-statistics}.
In short, when $T=9$, \tool\ achieves reasonably high accuracy. We interpret the
results as overall encouraging and reasonable, and we recommend using $T=9$ as
the default setup in practice.

\begin{table}[!thp]
  \centering
  \scriptsize
  \caption{Distribution of schemes which trigger outliers when $T=9$. \protect\\
  LC represents Leetcode and CNS represents CodesearchNet.}
  \vspace{-5pt}
  \label{tab:pass-distribution}
  \setlength{\tabcolsep}{2pt}
  \resizebox{1.00\linewidth}{!}{
    \begin{tabular}{l|c|c|c|c|c|c|c|c|c}
      \hline
      \textbf{Pass name} & \texttt{IRR} & \texttt{GRA_R} & \texttt{GRA_C}&  \texttt{REP_R} &  \texttt{REP_C}& \texttt{INI} & \texttt{RTF}& \texttt{REL_R} & \texttt{REL_C} \\
      \hline
      \textbf{\#variants on LC}            & 253& 598 & 594& 601& 600& 601& 38& 505& 506 \\
      \textbf{\#variants on CNS}           & 153& 2297& 2297& 2207& 2207& 2295& 314& 1914& 1918 \\
      \textbf{outliers on LC (\%)}          & 3.51& 17.03& 17.57& 11.35& 8.38& 12.7& 0.54& 17.03& 11.89 \\
      \textbf{outliers on CNS (\%)}         & 0.27& 15.41& 18.11& 11.62& 14.59& 11.62& 1.62& 14.32& 12.43 \\
      \hline
    \end{tabular}
  }
  \vspace{-4pt}
\end{table}

\parh{Potency of PSC Transformations.}~We measure the potency of all
nine PSC transformations about their contributions to uncovering outliers. 
For both LeetCode and CodeSearchNet, we report the distribution of generated
variants and triggered outliers in \T~\ref{tab:pass-distribution}. Due to
limited space, we report the average distribution across all code completion
systems; see~\cite{snapshot} for each system.
Overall, all PSC transformations reasonably facilitate uncovering outliers, even
though they are performed toward different levels of program hierarchical
representation (e.g., identifier level or basic block level). In particular,
nearly all schemes, except ``IRR'' and ``RTF'', have comparable contributions. IRR
has a relatively minor application scope, given that IRR looks for specific
arithmetic operations like ``+='' that may not be pervasively used. Similarly,
RTF requires the existence of an \texttt{if} branch with a relatively simple
condition, which may not be available in our test cases. As a consequence, these
two schemes result in less amount of uncovered outliers, as shown in the last two
rows in \T\ref{tab:pass-distribution}.

\begin{figure*}[!htbp]
  \centering 
  \includegraphics[width=1.0\linewidth]{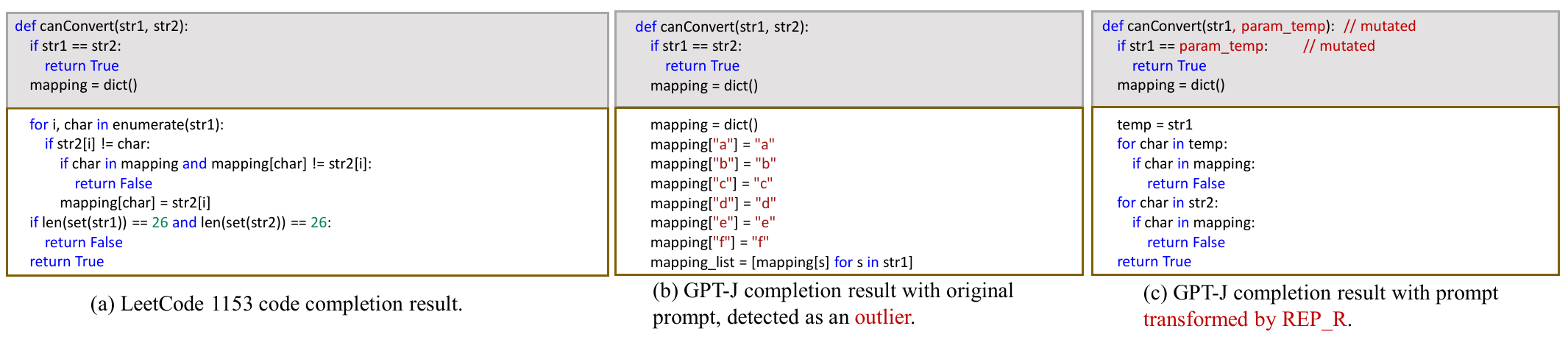}
  \vspace{-25pt}
  \caption{Outlier case study. The code completion output of the seed prompt is
  largely deviated, whereas a mutated prompt results in much better code
  completion output. The ``completion result'' (code in white) in (a) means
  ground truth.}
  \label{fig:outlier1}
\end{figure*}

\begin{figure*}[!htbp]
  \centering 
  \includegraphics[width=1.00\linewidth]{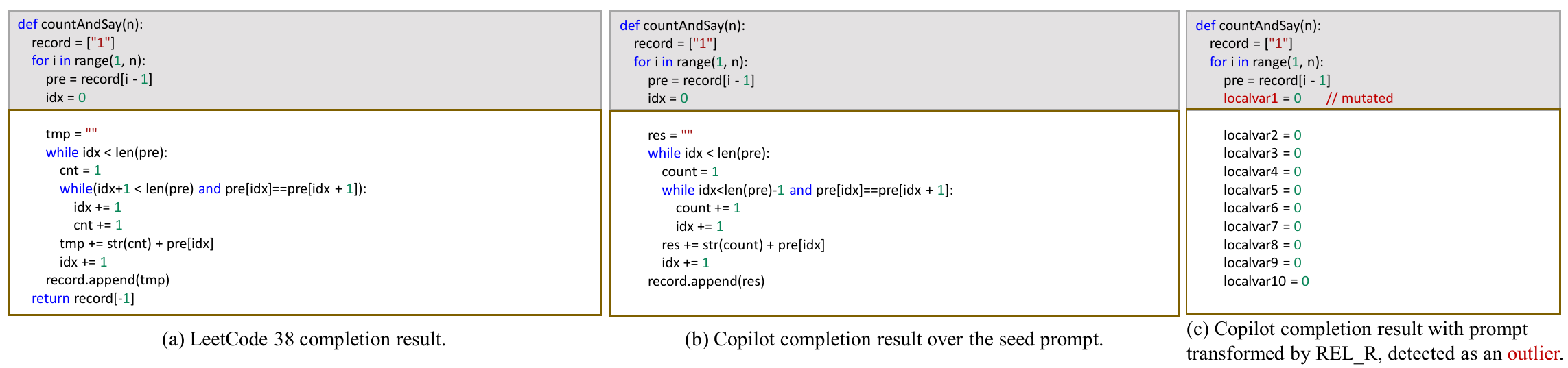}
  \vspace{-25pt}
  \caption{Outlier case study. The code completion output of a mutated prompt is
  largely deviated, whereas a the seed prompt results in much better code
  completion output. The ``completion result'' (code in white) in (a) means
  ground truth.}
  \label{fig:outlier2}
\end{figure*}

\parh{Case Studies.}~\T~\ref{tab:results} illustrates that the inconsistency
bugs in code completion systems are pervasive, and numerous defects can be found
even for highly permissive consistency thresholds. We present two representative
cases in \F~\ref{fig:outlier1} and \F~\ref{fig:outlier2}. 
\F~\ref{fig:outlier1} presents a case such that the completion output of the
seed prompt largely deviates from the ground truth. When only given the
completion output in \F~\hyperref[fig:outlier1]{5(b)}, it is inherently
challenging to decide if the deviation between completion outputs in
\F~\hyperref[fig:outlier1]{5(a)} and \F~\hyperref[fig:outlier1]{5(b)} is due to
model capacity or bugs. Nevertheless, when referring to
\F~\hyperref[fig:outlier1]{5(c)}, it becomes evident that the tested code
completion system, GPT-J, is \textit{capable} of generating high-quality
completion outputs with closer structure consistency to the ground truth. We
only tweak a little on the parameter name for the prompt in
\F~\hyperref[fig:outlier1]{5(c)} compared to the seed. Thus, it should be
accurate to consider \F~\hyperref[fig:outlier1]{5(b)} as code completion bug,
which may be likely repairable.

\F~\hyperref[fig:outlier2]{6(a)} presents another case, such that the code
completion output of the seed prompt appears to be highly similar to the ground
truth in \F~\hyperref[fig:outlier2]{6(b)}. In contrast, when applying the REL\_R
scheme to mutate a local variable's name, the code completion output of a
mutated prompt (as in \F~\hyperref[fig:outlier2]{6(c)}) becomes largely distinct
from the ground truth, which clearly denotes a bug in the code completion
system.

\parh{Time \& Cost.}~We employ a GPU server for running the involved models
locally. The server has an Intel Xeon Platinum 8276 CPU, 256 GB memory, and 4
NVIDIA A100 GPUs. Although processing time is generally not a concern, we record
and report that it takes 35 GPU seconds to finish completing one prompt on
average. For each seed, \tool\ takes 2.7 CPU seconds to generate nine mutated
prompts and about 82 GPU seconds for code completion systems to infer and obtain
the results in parallel. The outlier detection and enhancement (discussed in
\textbf{RQ3}) steps take about 1.1 CPU seconds. Overall, we admit that
generating prediction from many mutated prompts and ``averaging'' them will take
time, compromising interactive response time. Nevertheless, \textbf{RQ3}
illustrates that code completion outputs are of much higher quality.
Moreover, for commercial models, \T~\ref{tab:results} has revealed that roughly
10.86\% percent of its code completion outputs are spurious even when $T=9$. In
other words, we interpret that about 10.86\% percent of the code completion
outputs are highly confusing to the users and are thus ``wasted.'' This may
indicate an undesirable situation and potential \textit{financial loss}, given
that modern cloud-based code completion systems may feature a ``pay-as-you-go''
mode, where users are charged based on how many queries they send to the
services.

\begin{tcolorbox}[size=small]
\textbf{Answer to RQ2}: \tool\ identifies numerous defects when being used to
test (commercial) code completion systems, despite the varying thresholds used
in deciding outliers. We recommend configuring $T=9$ as a presumably proper
threshold (with the highest TP rates) in usage.
\end{tcolorbox}
\vspace{-2pt}

\subsection{RQ3: Enhancement Effectiveness}
\label{subsec:rq3}

To answer \textbf{RQ3}, we first present the improved accuracy of code
completion systems. Then, we assess the potency of each PSC scheme about
their contributions to enhancement. Lastly, we conduct a human evaluation to
verify the effectiveness of \tool in enhancing code completion tools.

\begin{table*}[!thpb]
  \centering
  \caption{Enhancement result When $T=9$. Values denote improvement ratios
  against baseline, instead of absolute improvement amounts.}
  \vspace{-5pt}
  \label{tab:enhancement}
  \setlength{\tabcolsep}{2pt}
  \resizebox{1.00\linewidth}{!}{
    \begin{tabular}{l|c|c|c|c|c|c|c|c|c}
      \hline
                              & \multicolumn{8}{c|}{\textbf{Code Completion Systems}} & \multirow{2}{*}{\textbf{Avg. Enhancement}} \\\cline{2-9}
                              & Copilot & CodeParrot & CodeParrot-small & GPT-J & GPT-NEO-13B & GPT-NEO-125M & Codegen-2B-mono& Codegen-6B-mono &  \\\cline{2-9}
      \hline
      \textbf{BLEU (\%)}            & 21.22 + 33.98   & 59.96 + 50.21  & 38.31 + 42.85   &36.14 + 46.9 & 40.58 + 50.04 & 52.74 + 54.71    & 39.81 + 22.96 &32.84 + 21.02  & 40.20 + 40.33\\
      \textbf{Edit Sim (\%)}        & 4.77 + 38.77   & 56.74 + 64.45  & 53.41 + 65.74    &87.93 + 80.16 & 102.3 + 82.8 & 100.69 + 110.84  & 52.81 + 72.93 & 37.48 + 67.17 & 62.01 + 72.85\\
      \hline
    \end{tabular}
  }
  \vspace{-5pt}
\end{table*}

\parh{Improved Accuracy.}~We report the improved accuracy with respect to both
Levenshtein edit similarity \cite{Fuzzywuzzy} and BLEU score
\cite{Papineni02bleu} in \T~\ref{tab:enhancement} when $T=9$. Note that both
metrics are commonly used in relevant research to assess the performance of LLMs
\cite{guo2020graphcodebert,guo2022unixcoder}; a higher edit similarity score or
BLEU score between the ``ground truth'' and the completion output indicates
better performance. To clarify, \T~\ref{tab:enhancement} reports the enhancement
ratio. For instance, when assessing Copilot against the LeetCode dataset, let
the edit similarity or BLEU score be $s$. We compute the enhancement ratio as $r
= \frac{s' - s}{s}$, where $s'$ is the edit similarity/BLEU score after
enhancement. Similar to \textbf{RQ2}, ``21.22 + 33.98'' in the first Copilot
cell means that the relative gain of average BLEU score on LeetCode dataset is
21.22\% and on CodeSearchNet dataset is 33.98\%. 

\parh{Cost.}~As noted in \textbf{Time \& Cost} paragraph in \S~\ref{subsec:rq2},
processing all prompts mutated from a seed takes about 82 GPU seconds on
average, while the cost of outlier detection and enhancement selection is about
1.1 CPU seconds. Therefore, we estimate that an acceptable cost is incurred for
a substantial improvement. Note, however, that commercial code completion
systems may have a ``pay-per-query'' mode, in which each query is charged.
Therefore, it may be preferable for the service provider (instead of users) to
deploy \tool.

\parh{Potency.}~Let completion output $o_{i}$ be the output used for
enhancement, and $o_{i}$ is generated using the prompt mutated from PSC
transformation $t_i$. We deem $t_i$ under this circumstance as the ``optimal''
transformation that contributes to the code completion enhancement. 
For both LeetCode and CodeSearchNet, we report the distribution of different
transformations selected as ``optimal'' in \T\ref{tab:rq3-pass-distribution}. We
find all PSC transformations are effectively served as ``optimal'' in both
datasets. As clarified in the ``potency'' evaluation in \textbf{RQ2}, ``IRR''
and ``RTF'' have smaller application scope and fewer variants, which indicates
their relatively lower proportion of being ``optimal'' on both datasets.
However, an exception is that ``IRR'' contributes disproportionately well when
using the LeetCode dataset. With manual inspection, we find that the programs in
LeetCode are more likely to contain arithmetic operations such as ``+=''
compared to programs in CodeSearchNet. In general, we interpret the evaluation
results are highly encouraging, showing that almost all the designed schemes
manifest high applicability and effectiveness in enhancing different models.

\begin{table}[!thp]
  \centering
  \scriptsize
  \caption{Distribution of ``optimal'' PSC transformations contributing to the
  output enhancement when $T$ is 9. \protect\\ LC represents Leetcode and CNS represents CodesearchNet}
  \vspace{-5pt}
  \label{tab:rq3-pass-distribution}
  \setlength{\tabcolsep}{2pt}
  \resizebox{1.00\linewidth}{!}{
    \begin{tabular}{l|c|c|c|c|c|c|c|c|c|c}
      \hline
      \textbf{Pass name} & \texttt{IRR} & \texttt{GRA_R} & \texttt{GRA_C}&  \texttt{REP_R} &  \texttt{REP_C}& \texttt{INI} & \texttt{RTF}& \texttt{REL_R} & \texttt{REL_C} & Total \\
      \hline
      \textbf{ BLEU on LC (\%)}              & 7.12& 2.65& 3.14& 2.79& 4.78& 10.80& 0.27& 4.76& 3.89 &40.20\\
      \textbf{ BLEU on CNS (\%)}             & 0.21& 4.21& 3.47& 4.59& 5.88& 5.89& 0.59& 5.57 & 9.92 &40.33\\
      \textbf{ Edit Sim on LC (\%)}          & 5.32& 4.81& 3.60& 9.03& 8.13& 12.59& 1.37& 8.46 & 8.70 &62.01 \\
      \textbf{ Edit Sim on CNS (\%)}         & 0.20& 5.89& 6.15& 8.66& 8.27& 13.40& 1.38& 14.12& 14.77& 72.85 \\
      \hline
    \end{tabular}
  }
  \vspace{-4pt}
\end{table}

\parh{Human Study.}~We further conduct human evaluation to inspect the quality
of the enhanced completion outputs. We randomly select 60 samples and
create an online questionnaire for them. We invite twelve
experts, including four industrial developers and eight academy researchers with
expertise in LLMs, as the participants. We provide two completion outputs for
each seed program without specifying whether they are the original or the
\tool's enhanced outputs. On a scale from 1 to 5, the participants are asked to
provide a score (1 for a completely not satisfying output and 5 for a fully
satisfying output). 

To ensure that all participants correctly understood the meaning of the
questionnaire, we prepare and insert five sanity-check (SC) test items randomly
into the questionnaire, and only participants who answer all SC correctly are
considered legitimate. We evenly assigned 30 real samples with five SC to each
participant, and ensured that each selected sample was examined by six
participants. All participants passed the sanity check and the average
completion time was 45 minutes.

The human evaluation results show that the average score for the original
outputs is 2.3188, whereas for \tool's enhanced outputs, the average score is
3.1565, representing a relative gain of 36.12\%. We analyzed all their answers.
Respondents believe that for 2.2\% of the cases, the enhanced completion outputs
look worse than the original outputs, while the remaining 97.8\% treat the
enhanced outputs equal to or better than the original ones. The Fleiss' Kappa
score~\cite{fleiss1971measuring} for the questionnaire is 0.94, which can be
interpreted as ``almost perfect agreement'' between human participants. These
findings indicate \tool's effectiveness in enhancing code completion outputs.

\vspace{-2pt}
\begin{tcolorbox}[size=small]
\textbf{Answer to RQ3}: \tool\ improve the outputs of different code completion
systems. We also find that instead of one or a few PSC transformations that
significantly enhance code completion, all schemes are shown as effective.
\end{tcolorbox}
\vspace{-2pt}
\section{Discussion}
\label{sec:discussion}

\noindent \textbf{Limitations and Threats to Validity.}~We now discuss the
validity and limitations of this work. In this research, \textit{construct
validity} denotes the degree to which our metrics reflect the correctness of
code completion systems. Overall, we conduct automatic testing and manual
inspection to study the outputs of de facto code completion systems. Hence,
while this practical approach detects their defects and reveals chances of
enhancing their outputs, a possible threat is that our testing approach
cannot guarantee the correctness of code completion systems. We clarify that our
work roots the same assumption as previous testing works that aim to detect
flaws of deep learning-based applications with dynamic testing rather than
verification.

We check code completion outputs by comparing a set of outputs that are supposed
to be (visually) consistent. The evaluation shows that the focus of structural
consistency effectively unveils a large number of defects. However, a possible
threat is that defects can be neglected in the completion output, in case all
outputs share aligned yet erroneous code patterns. We deem this a general and
well-known hurdle for invariant property-based testing techniques. We leave
exploring solutions to this challenge for future work.

Besides, the potential threat exists that the proposed testing and enhancing 
framework, \tool, may not adapt to other types of code completion systems.
Nevertheless, we mitigate this threat to \textit{external validity} by designing
a system and algorithm independent approach. As a result, our approach is
anticipated to apply to other settings outside the current scope. We
believe the proposed technique is general, and we give further discussions
regarding other settings in this section. 

\parh{``Natural-Looking'' Mutations.}~\tool\ designs nine PSC transformations to
mutate prompts. Nevertheless, one may question if all of these mutations are
common in real-world programming. Holistically, we agree that some
transformations, e.g., renaming variables, may lead to potentially rare names,
e.g., using \texttt{LocalVar1} as a variable name; as illustrated in under the
\texttt{REL_R} scheme.
Nevertheless, PSC transformations is intentionally designed as
\textit{lightweight}, for the seek of easing the follow-up outlier detection. We
find that the present transformations are sufficient to expose many defects and
harvest research insights. 
More importantly, we focus on designing \textit{structure-preserving}
transformations. Too aggressive transformations may easily break the structure
consistency of mutated prompts, making our testing oracle inapplicable.
Thus, while we also consider the specific program context (as noted in
\T~\ref{tab:psc}) to enhance the ``realism'' of mutated prompts, realism is not
our highest priority. 

Overall, generating natural-looking, realistic inputs to test deep learning
systems is inherently challenging. For instance, some computer vision (CV)
related testing may need to use expensive generative models like GAN to generate
more natural-looking images and test auto-driving
systems~\cite{zhang2018deeproad}. It is unclear if those methods fit our
scenarios.
As a future work, we plan to study using advanced methods (e.g., LLMs) to
generate mutated test inputs that are both ``natural-looking'' and
structure-preserving. The cost may be a primary concern, as we generated around
20K mutated test inputs in the evaluation.

\parh{Cross Comparison of Code Completion Outputs.}~Overall, \tool\
\textit{individually} tests each code completion system. The proposed approach
constitutes program property-based testing (or metamorphic testing). 
With this regard, careful readers may wonder about the feasibility of conducting
\textit{differential testing} by processing the same prompt with different code
completion systems and differentiating their outputs. However, we note that the
code completion outputs can have drastically different representations since
different code completion systems have different model training data and LLM
model capacity. For instance, Copilot is seen to produce a large chunk of code
snippets (with multiple statements), whereas some other well-known systems are
prone to giving more succinct outputs for the same input prompts. Our
preliminary exploration also shows that they manifest different tactics and
translation templates in code generation. Thus, the similarity among the code
completion outputs is deemed as \textit{low} across different code completion
systems. Overall, we leave it as one future work to explore practical methods to
perform cross comparison, for instance, by extracting specific
``semantics-level'' signatures or regulating their output code patterns first.

\parh{Other Settings.}~\tool\ targets Python code completion, one challenging
and popular task in software engineering. We believe it is feasible to migrate
\tool\ to test code completion systems of other languages like C and Java. While
extending \tool\ to handle other languages demands new parsers and
re-implementing PSC schemes, we expect that the key technical pipeline,
including mutation, outlier detection, and enhancing, are language
\textit{independent}. We thus believe migration is an engineering task rather
than open-ended research problems.

\section{Related Work}
\label{sec:related}

\parh{Testing Neural Models.}~Many works have applied testing methods to neural
models~\cite{zhang2018deeproad, tian2018deeptest, pei2017deepxplore,
zhou2018deepbillboard, odena2018tensorfuzz, wang2019adversarial,
dwarakanath2018identifying,
nakajima2019generating,yuan2022unveiling,pang2021mdpfuzzer,yuan2021perception}.
The tested neural models include computer vision tasks like image
classification, object
detection~\cite{wang2020metamorphic,shao2021testing,tian2021extent},
auto-driving~\cite{zhang2018deeproad,zhou2018deepbillboard}, as well as natural
language processing tasks like sentiment
analysis~\cite{ribeiro2020beyond,ma2020metamorphic,galhotra2017fairness,udeshi2018automated},
question answering~\cite{10.1109/ASE51524.2021.9678670}, machine
translation~\cite{he2020structure,he2021testing,gupta2020machine,sun2020automatic},
and specific properties like fairness~\cite{chen2022fairness}. Recent advances
in machine translation testing inspire \tool. However, \tool\ addresses
domain-specific challenges in mutating prompts, and for the first time, provides
a systematic framework for testing code completion systems in black-box
settings. 

\parh{Neural Code Comprehension.}~Besides code completion task (the focus of
\tool), neural models are used in other code comprehension
tasks~\cite{bielik2020adversarial,yefet2020adversarial,
rabin2021generalizability,rabin2021understanding,pour2021search,
suneja2021probing,henke2022semantic}. For instance, function naming decides the
function name by summarizing the function
body~\cite{alon2018code2seq,alon2019code2vec,zugner2021language}. Often, code
paths on the function ASTs are extracted for embedding and name prediction.
Given the large volume of available paths, optimization schemes like attention
are used to speed up the processing. Code classification and code search are two
popular tasks. Recent methods learn from structural information (AST and CFG)
for code
classification~\cite{mou2016convolutional,lu2021codexglue,wang2021codet5} and
code
search~\cite{DBLP:conf/sigsoft/CambroneroLKS019,DBLP:journals/nn/GuLGWZXL21}.
Tree-based convolutional neural networks and graphics neural networks are
leveraged in these
tasks~\cite{wang2020combining,DBLP:journals/corr/abs-2111-02671,wang2022sem2vec,wang2022enhancing}.
Software security applications have been built based on neural code
comprehension, including plagiarism
detection~\cite{lei2022deep,DBLP:conf/icse/Li0WW0NW22}, malware
clustering~\cite{kalash2018malware}, software component
analysis~\cite{zhan2021atvhunter,woo2021centris,yu2020codecmr}, and
vulnerability detection~\cite{li2018vuldeepecker,zhou2019devign}. \tool\ differs
from these code comprehension tasks in that it focuses on testing code
completion systems. While some recent
works~\cite{bielik2020adversarial,pour2021search,henke2022semantic} treat neural
models as a ``white-box'' for retraining and robustness augmentation on
relatively simple DNNs, \tool\ delivers black-box repairing over LLMs.

\section{Conclusion}
\label{sec:conclusion}

We offer \tool, a testing and repairing tool for code completion systems. \tool\
uncovers thousands of defects, and its enhancement improves their output quality
largely. This work may serve as a roadmap for researchers and users interested
in using and improving code completion systems.

\section*{Acknowledgement}
We thank anonymous reviewers for their valuable feedback. HKUST authors are
supported in part by a RGC ECS grant under the contract 26206520.

\bibliographystyle{plain}
\bibliography{bib/ref,bib/analysis,bib/decompiler,bib/testing-cv,bib/cv}

\end{document}